\documentclass[fleqn,100pt]{wlscirep}
\usepackage{amsmath,amssymb,wasysym}
\usepackage{xcolor}

\usepackage{siunitx}
\usepackage{lineno}
\title{Thickness-dependent magnetic order in CrI$_3$ single crystals}

\author[1*]{Yu Liu}
\author[1]{Lijun Wu}
\author[2]{Xiao Tong}
\author[1]{Jun Li}
\author[1]{Jing Tao}
\author[1]{Yimei Zhu}
\author[1*]{C. Petrovic}
\affil[1]{Condensed Matter Physics and Materials Science Department, Brookhaven National Laboratory, Upton, New York 11973, USA}
\affil[2]{Center for Functional Nanomaterials, Brookhaven National Laboratory, Upton, New York 11973, USA}
\affil[*]{yuliu@bnl.gov and petrovic@bnl.gov}

\begin{abstract}
Two-dimensional (2D) materials with intrinsic ferromagnetism provide unique opportunity to engineer new functionalities in nano-spintronics. One such material is CrI$_3$, showing long-range magnetic order in monolayer with the Curie temperature ($T_c$) of 45 K. Here we study detailed evolution of magnetic transition and magnetic critical properties in response to systematic reduction in crystal thickness down to 50 nm. Bulk $T_c$ of 61 K is gradually suppressed to 57 K, however, the satellite transition at $T^*$ = 45 K is observed layer-independent at fixed magnetic field of 1 kOe. The origin of $T^*$ is proposed to be a crossover from pinning to depinning of magnetic domain walls. The reduction of thickness facilitates a field-driven metamagnetic transition around 20 kOe with out-of-plane field, in contrast to the continuous changes with in-plane field. The critical analysis around $T_c$ elucidates the mean-field type interactions in microscale-thick CrI$_3$.
\end{abstract}
\begin{document}

\flushbottom
\maketitle
%
%
\thispagestyle{empty}

\section*{Introduction}

Layered materials, when thinned to atomic limits in 2D exhibit novel properties, different from the bulk counterpart. Recent discoveries of intrinsic 2D ferromagnetism in atomically thin CrI$_3$ and Cr$_2$Ge$_2$Te$_6$ open up new opportunities for studying fundamental 2D magnetism and show great potential in spintronic applications.\cite{McGuire,Huang,Gong} CrI$_3$ and Cr$_2$Ge$_2$Te$_6$ are ferromagnetic (FM) below $T_c$ $\sim$ 61 K in bulk.\cite{Carteaux2,Casto,Zhang} First-principle calculations predict a robust 2D ferromagnetism with $T_c$ $\sim$ 57.2 or 106 K in monolayer Cr$_2$Ge$_2$Te$_6$,\cite{Li,Sivadas} However, the scanning magneto-optic Kerr microscopy experiment shows that the $T_c$ monotonically decreases with decreasing thickness, showing 30 K in bilayer and the absence of $T_{c}$ in monolayer.\cite{Gong}

In bulk CrI$_3$, the Cr atoms in each layer form a honeycomb structure, and each Cr atom is surrounded by six I atoms in an octahedral coordination [Fig. 1(a)]. In contrast to Cr$_2$Ge$_2$Te$_6$, ferromagnetism in CrI$_3$ persists in monolayer with $T_c$ of 45 K.\cite{Huang} Intriguingly, the magnetism in CrI$_3$ is layer-dependent, from FM in monolayer to antiferromagnetic (AFM) in bilayer, and back to FM in trilayer, providing great opportunities for designing magneto-optoelectronic devices.\cite{Huang} The monolayer CrI$_3$ can be described by the Ising model.\cite{Huang,Griffiths} Remarkably, the electrostatic doping can modify the saturation magnetization, the coercive force, and the $T_c$ in monolayer as well as change the interlayer magnetic order in bilayer CrI$_3$.\cite{Jiang,Huang1} In addition, giant tunneling magnetoresistance is observed in few-layer CrI$_3$, exhibiting multiple states as a function of magnetic field.\cite{SongTC,Klein,WangZ,Kim}  This highlights CrI$_3$ as a potential magnetic tunnel barrier for van der Waals heterostructure spintronic devices and also shows demand for magnetic order investigation at all length scales from bulk to monolayer.

Here we report a detailed study of thickness-dependent magnetic transition and critical behavior in CrI$_3$. An additional satellite transition at $T^*$ was observed just below $T_c$. The value of $T^*$ is found thickness-independent at fixed magnetic field, in contrast to gradually suppressed $T_c$ with reduction in thickness. Thin CrI$_3$ crystals show an increase in coercive field and distinct field-driven metamagnetic transition around 20 kOe with out-of-plane field but not in-plane field. The critical behavior suggests the long-range mean-field type interactions in mesoscale-thick CrI$_3$ crystals.

\section*{Methods}

Bulk CrI$_3$ single crystals were fabricated by the chemical vapor transport method and characterized as described previously.\cite{LIUYU} A series of thicknesses were obtained step-by-step mechanical exfoliating bulk crystal down to nanometer scale [Figs. 1(b) and 1(c)]. Considering the reactivity of thin CrI$_3$ flakes, all the samples were prepared in an argon-filled glove-box and protected using scotch tape on both sides when transferring for magnetization measurement. The dc magnetization was measured in Quantum Design MPMS-XL5 system. An optical microscopy equipped with 100$\times$ objective lens in Witec alpha 300 confocal Raman microscope was used for imaging the cross-section of S2 sample to determine its thickness. Transmission electron microscopy (TEM) sample of S3 was prepared by putting the exfoliated S3 sample in Cu grid. The thickness ($t$) of S3 sample was measured by the low loss EELS spectrum in JEM-ARM200F microscope [Fig. 1(c)], $t = ln(I_t/I_0) \times \lambda$, where $I_0$ and $I_t$ are integrated EELS signal intensity under the zero-loss peak and the total spectrum, respectively; $\lambda$ is a total mean free path for all inelastic scattering which can be calculated based on the composition of the sample.\cite{Egerton}

\section*{Results and Discussion}

\begin{figure}
\centerline{\includegraphics[scale=1]{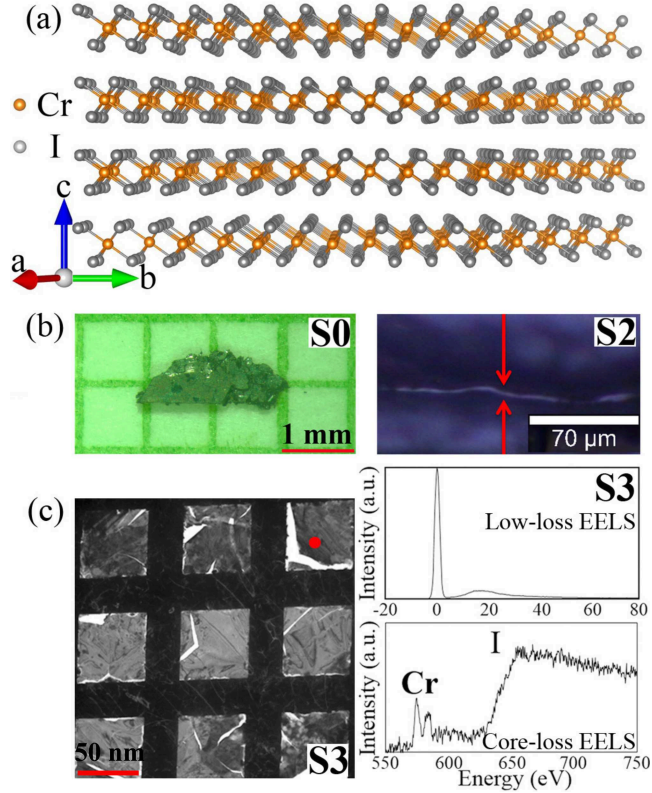}}
\caption{(Color online). (a) Low temperature structure of CrI$_3$ in the $R\bar{3}$ space group. (b) Optical microscopy of bulk S0 sample (thickness $\sim100\mu$m) and cross-section of S2 sample (thickness $\sim2\mu$m). (c) Transmission electron microscopy (TEM) image of S3 sample in Cu grid along with the low-loss and core-loss electron energy loss spectra (EELS) taken from top right area marked by red circle. The thickness of S3 was calculated from the low-loss EELS based on log-ratio method in Digital Micrograph program to be 44 nm. The average thickness for many areas measured is 49$\pm$8 nm.}
\label{thickness}
\end{figure}

\begin{figure}
\centerline{\includegraphics[scale=0.9]{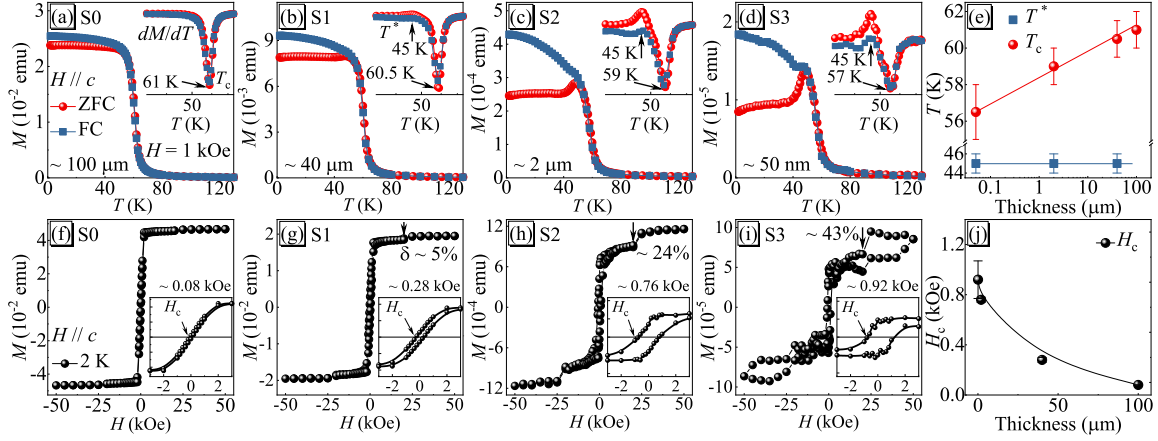}}
\caption{(Color online). (a-d) Temperature dependence of magnetization for CrI$_3$ at various crystal thicknesses measured with $H // c$ at 1 kOe. Insets show the first derivative of the $M(T)$ curves. (e) Thickness-dependent $T^*$ and $T_c$. (f-i) Isothermal magnetization at $T$ = 2 K for corresponding S0-S3 samples. Insets show the enlarged low field part and the derived coercive field $H_c$. (j) Thickness dependence of $H_c$.}
\label{MTH}
\end{figure}

Figures 2(a)-2(d) present the temperature-dependent magnetization $M(T)$ with zero-field cooling (ZFC) and field-cooling (FC) modes measured in out-of-plane field $H$ = 1 kOe for the samples (S0-S3) with typical thickness from 100 $\mu$m to 50 nm. There is an apparent increase in $M(T)$, in line with the paramagnetic (PM) to FM transition.\cite{McGuire} The $T_c$ can be determined from the minima of the first derivative of the $M(T)$ curves [insets in Figs. 2(a)-2(d)]. Given the large magnetocrystalline anisotropy in CrI$_3$,\cite{Richter} the divergence of ZFC and FC curves is observed below $T_c$ in $H$ = 1 kOe. With subsequent reduction of crystal thickness this discrepancy tends to be larger, indicating an increasing anisotropy in thin crystals. Intriguingly, an additional satellite transition $T^*$ just below $T_c$ is observed. A similar phenomenon was also observed in Fe$_3$GeTe$_2$, arising from the emergence of antiparallel spin arrangement along the $c$ axis between different Fe$_3$Ge layers.\cite{Yi} This satellite transition $T^*$ is defined by the maxima of the first derivative of the $M(T)$ curves [insets in Figs. 2(b)-2(d)], which is more apparent in thinner crystal. The thickness dependence of $T_c$ and $T^*$ is summarized in Fig. 2(e). The $T_c$ shows linear decrease with logarithm of thickness. In contrast, the $T^*$ seems thickness-independent and shows a value of 45 K identical to the $T_c$ in monolayer measured by magneto-optic Kerr microscopy in $H$ = 1.5 kOe.\cite{Huang} The isothermal magnetization at $T$ = 2 K for the corresponding samples is shown in Figs. 2(f)-2(i). With $H//c$, the magnetization of bulk CrI$_3$ crystal saturates at a relatively low magnetic field $H$ $\sim$ 2 kOe and in line with the previous reports.\cite{McGuire,LIUYU} With reduction of thickness, there is an increase of coercive field $H_c$ as shown in insets of Figs. 2(f)-2(i), from about 80 Oe for S0 sample to 920 Oe for S3 sample, still identified as a relatively soft ferromagnet. It is interesting to note that an apparent magnetization jump around $H$ $\sim$ 20 kOe is observed in thin crystals. The calculated value $\delta = [M(25 kOe)-M(20 kOe)]/M(20 kOe)$ increases from 5 $\%$ for S1 to 43 $\%$ for S3. This field-driven metamagnetic transition is first identified by the step-like change of magnetization in the hysteresis curves, which is similar with the giant tunneling magnetoresistance step around the same field observed in few-layer devices.\cite{Kim} The tunneling magnetoresistance step in few-layer devices is attributed to spin filtering effect arising from collapse of interlayer AFM in thin CrI$_3$.\cite{SongTC,Klein,WangZ,Kim} Above this critical field, a fully spin polarized FM state is reached in S0-S2 samples, however, a large hysteresis is observed in S3 sample.

\begin{figure}
\centerline{\includegraphics[scale=0.8]{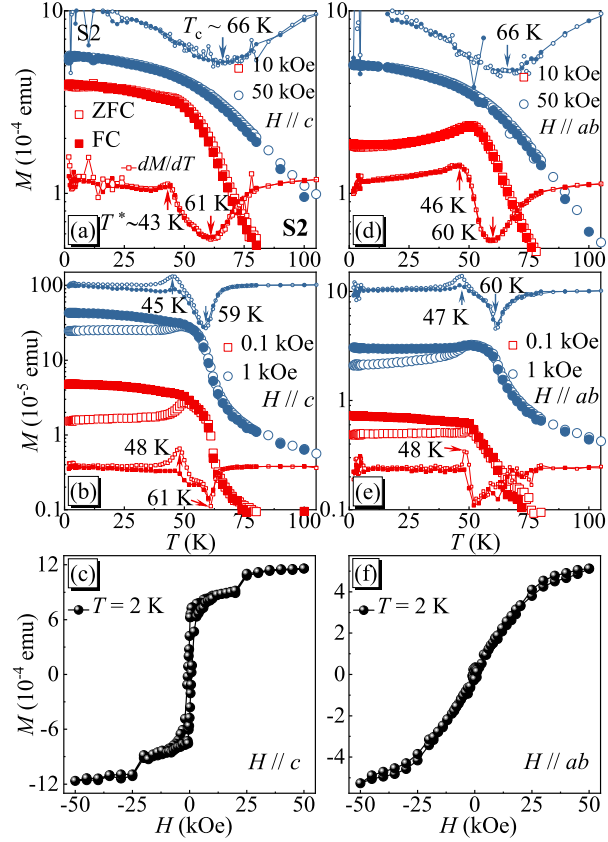}}
\caption{(Color online). (a,b) Temperature-dependent magnetization measured at various magnetic fields and (c) field-dependent magnetization measured at $T$ = 2 K for $H // c$. (d,e) Temperature-dependent magnetization measured at various magnetic fields and (f) field-dependent magnetization measured at $T$ = 2 K for $H // ab$.}
\label{MTH}
\end{figure}

To well understand the nature of magnetism in CrI$_3$ and its properties from bulk to monolayer, we then focus on the S2 sample with mesoscale thickness of 2 $\mu$m. Figure 3 shows the temperature- and field-dependent magnetization $M(T,H)$ with out-of-plane and in-plane fields, respectively. The $M(T)$ is nearly isotropic in $H$ = 50 kOe, whilst anisotropic magnetic response is observed in low fields. In $H$ = 10 kOe, a monotonic increase of $M(T)$ with decreasing temperature is observed for $H // c$ [Fig. 3(a)], however, the $M(T)$ decreases with decreasing temperature below 50 K for $H // ab$ [Fig. 3(d)]. Similar feature was also observed in bulk CrI$_3$ crystal, associated with its temperature dependent magnetocrystalline anisotropy.\cite{Richter,YULIU} The satellite transition $T^*$ reveals itself by the emergence of a weak kink in the $dM/dT$ curve with $H$ = 10 kOe. With decreasing field, the magnetic anisotropy results in larger divergence between ZFC and FC magnetization and a more abrupt satellite transition, as shown in Figs. 3(b) and 3(e). Both $T_c$ and $T^*$ are field-dependent but with an opposite tendency, i.e., with decreasing field the $T_c$ decreases whereas the $T^*$ gradually increases. The isothermal magnetization at 2 K increases discretely for out-of-plane field [Fig. 3(c)], but monotonically for in-plane field [Fig. 3(f)]. Although an A-type AFM ground state was proposed in few-layer devices,\cite{Kim} a FM ground state is still expected in S2 sample with thickness $\sim2\mu$m. Therefore, the field-driven metamagnetic transition is most likely caused by a movement or depinning of magnetic domains, rather than flipping the spins from an antiparallel to a parallel configuration. A crossover from pinning to depinning of magnetic domain walls is also proposed as the reason for the satellite $T^*$ kink. When the sample is cooled in zero field, the magnetic domains start to be pinned below the crossover temperature $T^*$. Small applied field at base temperature is insufficient to move the pinned domains. With increasing temperature, thermal fluctuations gradually weaken the pinning force and finally completely depin the domains above $T^*$. This also explains why the $T^*$ kink is more apparent in the ZFC curves. In the FC process the domains are always pinned with the effective FM moment aligned along the cooling field.

\begin{figure}
\centerline{\includegraphics[scale=1]{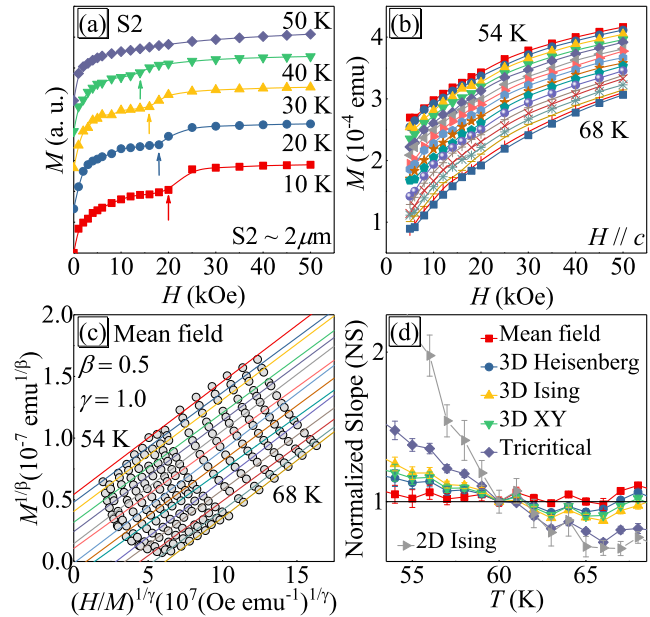}}
\caption{(Color online). (a,b) Initial isothermal magnetization at various temperatures with out-of-plane field. (c) The Arrott Plot of S2 sample with $\beta$ = 0.5 and $\gamma$ = 1.0. (d) Temperature-dependent $NS = S(T)/S(T_c)$ for indicated models.}
\label{models}
\end{figure}

Near a second-order phase transition, the free energy of a ferromagnet $F_m(M)=F_m(0)+\frac{1}{2}aM^2 +\frac{1}{4}bM^4 +...-HM$. By $\partial F_m(M)/\partial M = 0$, the equation of state is obtained as $H/M = a +bM^2$ for the behavior of magnetization near $T_c$. According to the scaling hypothesis, the spontaneous magnetization $M_s$ below $T_c$, the inverse initial susceptibility $H/M$ above $T_c$, and the $M(H)$ at $T_c$ can be characterized by a series of critical exponents:\cite{Stanley}
\begin{equation}
M_s (T) = M_0(-\varepsilon)^\beta, \varepsilon < 0, T < T_c,
\end{equation}
\begin{equation}
H/M(T) = (h_0/m_0)\varepsilon^\gamma, \varepsilon > 0, T > T_c,
\end{equation}
\begin{equation}
M = DH^{1/\delta}, \varepsilon = 0, T = T_c,
\end{equation}
where $M_0$, $h_0/m_0$, $D$ and $\varepsilon = (T-T_c)/T_c$ are the critical amplitudes and the reduced temperature, respectively.\cite{Fisher}

The isothermal magnetization curves with out-of-plane field are depicted in Figs. 4(a) and 4(b). The magnetization jump gradually moves to lower field and finally disappears at 50 K for S2. The Arrott plot of $M^2$ vs $H/M$ [Fig. 4(c)], with $\beta$ = 0.5 and $\gamma$ = 1.0,\cite{Arrott1,LinJ} gives a set of quasi-straight lines and the isotherm at $T_c$ pass through the origin, suggesting the mean-field type magnetic interactions in S2 sample. In addition, the positive slope of the straight lines indicates that it is a second-order transition based on the Banerjee$^\prime$s criterion.\cite{Banerjee}. The Arrot-Noaks equation of state gives a general modified Arrott plot $(H/M)^{1/\gamma} = a\varepsilon+bM^{1/\beta}$, where $a$ and $b$ are constants.\cite{Arrott2} Taking into consideration the three-dimensional (3D) critical behavior in bulk CrI$_3$ (S0 sample),\cite{LIUYU} and 2D Ising-like in monolayer,\cite{Huang} a set of possible exponents belonging to different models are used to build the modified Arrott plots.\cite{LeGuillou} Comparison the normalized slope ($NS = S(T)/S(T_c)$) with ideal value "1" can determine the most suitable model, where $S(T) = dM^{1/\beta}/d(H/M)^{1/\gamma}$. Figure 4(d) presents the $NS$ vs $T$ curves for indicated models, confirming that the mean-field model is the best and the 2D Ising model shows the largest deviation. That is to say, the critical behavior of mesoscale S2 sample is quite different when compared to bulk and monolayer.\cite{LIUYU,Huang}

\begin{figure}
\centerline{\includegraphics[scale=1]{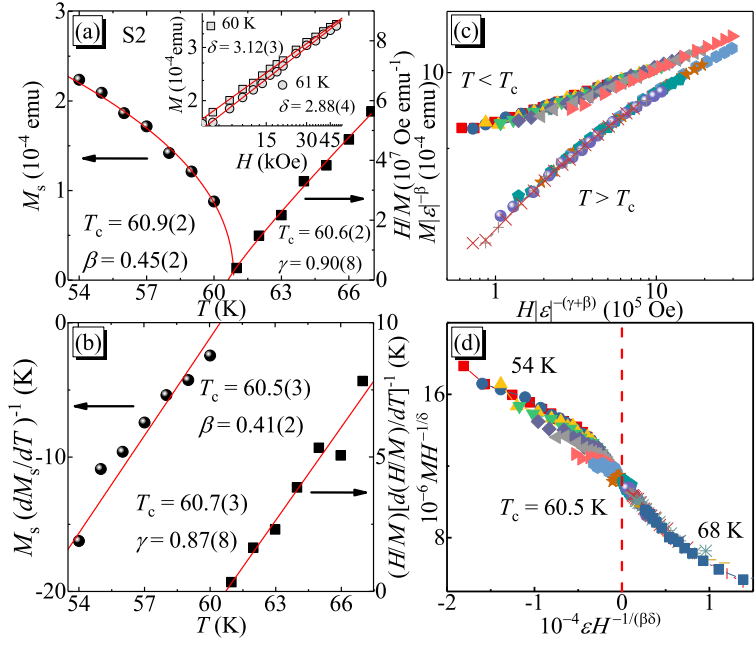}}
\caption{(Color online). (a) Temperature-dependent $M_s$ (left) and $H/M$ (right) with solid fitting curves. Inset shows the isotherms $M(H)$ in log-log scale near $T_c$ with linear fitting curves. (b) Temperature-dependent $M_s(dM_s/dT)^{-1}$ (left) and $(H/M)[d(H/M)/dT)]^{-1}$ (right) with linear fitting curves. (c) Scaling plot of $m$ vs $h$ with $m\equiv\varepsilon^{-\beta}M(H,\varepsilon)$ and $h\equiv\varepsilon^{-(\beta+\gamma)}H$, respectively. (d) $MH^{-1/\delta}$ vs $\varepsilon H^{-1/(\beta\delta)}$.}
\label{exponents}
\end{figure}

\begin{table}[ht]
\centering
\begin{tabular}{|l|l|l|l|l|l|l|l|}
\hline
  Thickness & Theoretical model & Reference & Technique & $T_c$ (K) & $\beta$ & $\gamma$ & $\delta$ \\
\hline
  S0 $\sim100\mu$m && \cite{LIUYU} & MAP & 60.45(3)/60.38(6) & 0.284(3) & 1.146(11) & 5.04(1) \\
  && \cite{LIUYU} & KFP & 60.05(13)/60.43(4) & 0.260(4) & 1.136(6) & 5.37(4) \\
  && \cite{LIUYU} & CI & 60 &   &   & 5.32(2) \\
  & 3D Heisenberg & \cite{Kaul} & Theory && 0.365 & 1.386 & 4.8 \\
  & 3D XY & \cite{Kaul} & Theory && 0.345 & 1.316 & 4.81 \\
  & 3D Ising & \cite{Kaul} & Theory && 0.325 & 1.24 & 4.82 \\
  & Tricritical mean field & \cite{LeGuillou} & Theory && 0.25 & 1.0 & 5\\
  S2 $\sim2\mu$m && This work & MAP & 60.9(2)/60.0(2) & 0.45(2) & 0.90(8) & 3.00(9) \\
  && This work & KFP & 60.5(3)/60.7(3)& 0.41(2) & 0.87(8) & 3.12(9) \\
  && This work & CI & 60 &   &   & 2.88(4) \\
  && This work & CI & 61 &   &   & 3.12(3) \\
  & Mean field & \cite{Stanley} & Theory && 0.5 & 1.0 & 3.0 \\
  monolayer && \cite{Huang} & MOK & 45 &  &  & \\
  & 2D Ising & \cite{LeGuillou} & Theory && 0.125 & 1.75 & 15 \\
\hline
\end{tabular}
\caption{\label{tab}The critical exponents of CrI$_3$ (S0 and S2) with different thickness obtained by various methods such as the modified Arrott plot (MAP), the Kouvel-Fisher plot (KFP), the critical isotherm (CI) and the magneto-optical Kerr (MOK) effect, in comparison with different theoretical models.}
\end{table}

The linearly extrapolated $M_s$ and $H/M$ are plotted from in Fig. 5(a). The solid lines are fitted lines according to Eqs. (1) and (2). The precise critical exponents $\beta = 0.45(2)$, with $T_c = 60.9(2)$ K, and $\gamma = 0.90(8)$, with $T_c = 60.6(2)$ K, are obtained, close to the parameters of mean-field model. According to Eq. (3), the $M(H)$ at $T_c$ should be a straight line in log-log scale with the slope of $1/\delta$ [inset in Fig. 5(a)]. Such fitting yields $\delta =$ 3.12(3) for $T$ = 60 K and 2.88(4) for $T$ = 61 K, respectively. The Kouvel-Fisher (KF) method can be used to double check the obtained critical exponents,
\begin{equation}
\frac{M_s(T)}{dM_s(T)/dT} = \frac{T-T_c}{\beta},
\end{equation}
\begin{equation}
\frac{H/M(T)}{d[H/M(T)]/dT} = \frac{T-T_c}{\gamma},
\end{equation}
where $M_s/(dM_s/dT)$ and $(H/M)/[d(H/M)/dT]$ are linear in $T$ and the slopes are $1/\beta$ and $1/\gamma$, respectively.\cite{Kouvel} Such fitting gives $\beta = 0.41(2)$, with $T_c = 60.5(3)$ K, and $\gamma = 0.87(8)$, with $T_c = 60.7(3)$ K, respectively, in line with the values obtained by the modified Arrott plot.

The reliability of obtained critical exponents $\beta$, $\gamma$, $\delta$, and $T_c$ can be further checked by a scaling analysis. Near the critical region, the magnetic equation of state can be expressed as
\begin{equation}
M(H,\varepsilon) = \varepsilon^\beta f_\pm(H/\varepsilon^{\beta+\gamma}),
\end{equation}
where $f_\pm$ is the regular function with $f_-$ for $T<T_c$, and $f_+$ for $T>T_c$, respectively. This equation can be further expressed as $m = f_\pm(h)$, where $m\equiv\varepsilon^{-\beta}M(H,\varepsilon)$ and $h\equiv\varepsilon^{-(\beta+\gamma)}H$ are the rescaled magnetization and magnetic field, respectively. Figure 5(c) presents the scaled $m$ vs $h$ in log-log scale, falling on two separate branches above and below $T_c$, confirming the true scaling treatment and the intrinsic critical values of $\beta$, $\gamma$, and $\delta$. The scaling equation of state can also be expressed as
\begin{equation}
\frac{H}{M^\delta} = k(\frac{\varepsilon}{H^{1/\beta}}),
\end{equation}
where the $MH^{-1/\delta}$ vs $\varepsilon H^{-1/(\beta\delta)}$ should collapse into a single curve with the $T_c$ locating at horizontal zero point [Fig. 5(d)].

The critical exponents of CrI$_3$ (S0 and S2) with different thickness are listed in Table I for comparison with the theoretical models.\cite{Huang,LIUYU,Stanley,Kaul,LeGuillou} For a 2D magnet, the value of critical exponent $\beta$ should be within a window $\sim$ $0.1 \leq \beta \leq 0.25$.\cite{Taroni} As we can see, the critical exponents of bulk CrI$_3$ crystal (S0) are between the values of theoretical tricritical mean-field and 3D Ising model, exhibiting a clear 3D critical phenomenon, in contrast to Cr$_2$(Si,Ge)$_2$Te$_6$ showing 2D Ising-like behavior.\cite{YuLiu,Lin} The monolayer CrI$_3$ is claimed as a 2D Ising-like ferromagnetism as a result of large magnetic anisotropy removing the Mermin-Wagner restriction.\cite{Huang,Mermin} However, the critical exponents of $\beta$, $\gamma$, and $\delta$ for mesoscale CrI$_3$ (S2) are well fit with the mean-field model (Table I), which is quite different from the behavior in bulk and monolayer. It should be noted that the precise value of $T_c$ for S2 almost equals to $T_c$ = 61 K for S0 based on the critical analysis, in contrast to the strong thickness dependence of $\beta$, $\gamma$, and $\delta$. we assume that the $T_c$ of 45 K in monolayer is most likely stabilized by the satellite transition ($T^*$ $\sim$ 45 K ) in bulk CrI$_3$, calling for further in-depth theoretical study.

\section*{Conclusion}

In summary, we have investigated the thickness-dependent magnetism in CrI$_3$ from bulk to 50 nm thickness. In addition to suppression of bulk $T_{c}$, we observe an additional satellite transition $T^*$ at 45 K which is independent of thickness and corresponds to the $T_c$ observed in monolayer. Analysis of critical behavior shows mean-field interactions in mesoscale-thick CrI$_3$, in contrast to that in bulk and monolayer, which is helpful to understand the nature magnetism in in CrI$_3$ and is of great importance for future magneto-optoelectronic devices.


\section*{Acknowledgements}

Work at Brookhaven National Laboratory is supported by the US DOE, Contract No. DE-SC0012704.

\section*{Author contributions statement}

C.P. initiated the study. Y.L. synthesized crystals and performed magnetization measurements. X.T. performed optical microscopy study. L.W., J.L, J.T., and Y.Z. contributed TEM and EELS measurements. Y.L. and C.P. organized and wrote the paper with input from all collaborators. This manuscript reflects the contribution and ideas of all authors.

\section*{Additional information}

Competing interests: The authors declare no competing interests.

\end{document}